\documentclass[preprint,prl,showpacs,aps]{revtex4-1}
\usepackage{graphicx}
\usepackage{pdfpages}
\begin{document}

\title{Hydrogen-bonded supramolecular assembly of dyes \\
at nanostructured solar cell interfaces}

\author{Christopher E. Patrick}
\author{Feliciano Giustino}
\affiliation{Department of Materials, University of Oxford, Parks Road, 
Oxford OX1 3PH, United Kingdom}

\begin{abstract}
We calculate from first principles the O$1s$ core-level shifts
for a variety of atomistic models of the interface between 
TiO$_2$ and the dye N3 found in dye-sensitized solar cells. 
A systematic comparison between our calculations
and published photoemission data shows that only
interface models incorporating hydrogen bonding 
between the dyes are compatible with experiment.
Based on our analysis we propose that at the TiO$_2$/N3 interface
the dyes are arranged in supramolecular assemblies. 
Our work opens a new direction in the modeling
of semiconductor/dye interfaces and bears on the design 
of more efficient nanostructured solar cells.
\end{abstract}

\date{\today}

\maketitle

Among promising low-cost photovoltaics, dye-sensitized solar
cells (DSCs) \cite{ORegan1991} based on nanostructured TiO$_2$ films sensitized with the
dye Ru(dcbpyH$_2$)$_2$(NCS)$_2$ (N3) have gained prominence
over the past two decades due to their relatively high energy
conversion efficiencies in excess of 10\% 
\cite{Nazeeruddin1993, Nazeeruddin1999, Bessho2009}.
In these devices the photocurrent is generated via ultrafast electron
transfer from the photoexcited dye to the 
nanostructured semiconductor
\cite{Duncan2007}.
Since the electron injection takes place within a sub-nanometer length scale,
the atomistic nature of the TiO$_2$/N3 interface
plays a critical role in the performance of DSCs \cite{Angelis2007}.
The dye N3 has four carboxylic acid groups \cite{sup_figs}.  
It is generally agreed that the adsorption of N3 onto
the anatase TiO$_2$ surface occurs through the anchoring of
one or more of these groups via the formation of Ti-O bonds
\cite{Nazeeruddin2003,Johansson2005,Rensmo1999,Schiffmann2010,Angelis2010}.
However the detailed atomic-scale structure of the TiO$_2$/N3 interface
remains controversial, and questions such as how many and which carboxylic groups 
participate in the bonding to the substrate are being actively debated
\cite{Nazeeruddin2003,Johansson2005,Rensmo1999,Schiffmann2010,Angelis2010}.

In this work we propose a new atomic-scale model of the TiO$_2$/N3 interface
by reverse-engineering measured X-ray photoemission spectra (XPS).
We first calculate from first-principles 
the O$1s$ core-level shifts for a variety of atomistic models of the
TiO$_2$/N3 interface. We then perform a quantitative comparison between 
our calculated core-level shifts and the XPS spectra of Ref.~\onlinecite{Johansson2005}.
Such comparison shows that only interface models 
which incorporate hydrogen bonding interactions
between the dyes are compatible with the measured spectra.
Based on our analysis we propose that at the
TiO$_2$/N3 interface the dyes are arranged in supramolecular 
hydrogen-bonded assemblies.

The existence of competing models of the atomic-scale structure
of the TiO$_2$/N3 interface illustrates the complexity of the problem.
Even in the case of an atomically perfect TiO$_2$ surface 
there exists a plethora of possible adsorption geometries \cite{Diebold2003}.
Previous computational studies have explored the potential 
energy landscape of one isolated N3 dye adsorbed on the TiO$_2$ surface, 
arguing in favor of specific models on the grounds of adsorption energy 
calculations \cite{Schiffmann2010, Angelis2010}.
The difficulties with this approach are that
(i) a thorough mapping of the total energy landscape 
is beyond current capabilities, and (ii) 
energetically favorable configurations may be
kinetically inaccessible during the fabrication of DSCs.
In order to avoid these difficulties from the outset we here follow
a completely different strategy and ask what is the atomic-scale model
of the TiO$_2$/N3 interface which best reproduces measured core-level
spectra. Our choice is motivated by the observation that core levels
are sensitive to the local bonding environment, and therefore carry the
signature of the atomistic interface structure.

We here consider the O$1s$ core-level shifts of TiO$_2$/N3 interfaces
reported in the XPS study of Ref~\citenum{Johansson2005}.
All our calculations are based on a generalized 
gradient approximation to density-functional theory,
and have been performed using the planewave pseudopotential
software package {\tt quantum ESPRESSO} \cite{quantumespresso}.
Core-level shifts are calculated using the theory 
developed in Refs.~\onlinecite{Pehlke1993,Pasquarello1996}.
A detailed description of our computational setup is given
as supplementary material \cite{sup_methods}.

\begin{figure}
\includegraphics{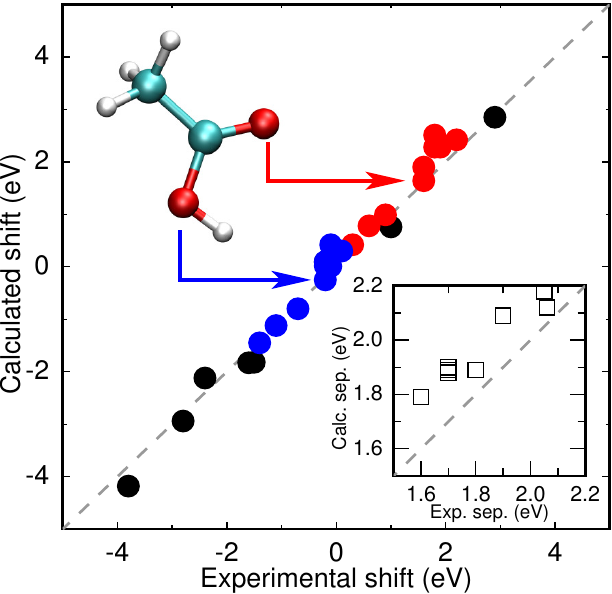}
\caption{\label{fig1}
Comparison of calculated O$1s$ core-level shifts of test molecules
\cite{sup_discussion}
with the experimental data of Refs.~\onlinecite{Jolly1984,McQuaide1988}.
The shifts are referenced to that of the H$_2$O molecule,
the binding energy increases towards negative energies.
Blue and red disks indicate the core-level shifts of molecules
containing carboxylic acid groups. The ball-and-stick
representation of the acetic acid shows the carbonyl and the hydroxyl O
atoms and the associated shifts (red and blue disks, respectively).
Inset: calculated splitting between the shifts of
hydroxyl and the carbonyl O atoms vs.~the experimental splitting.
The r.m.s.\ deviation is 0.17 eV (note the change of scale).
}
\end{figure}

\begin{figure*}
\includegraphics{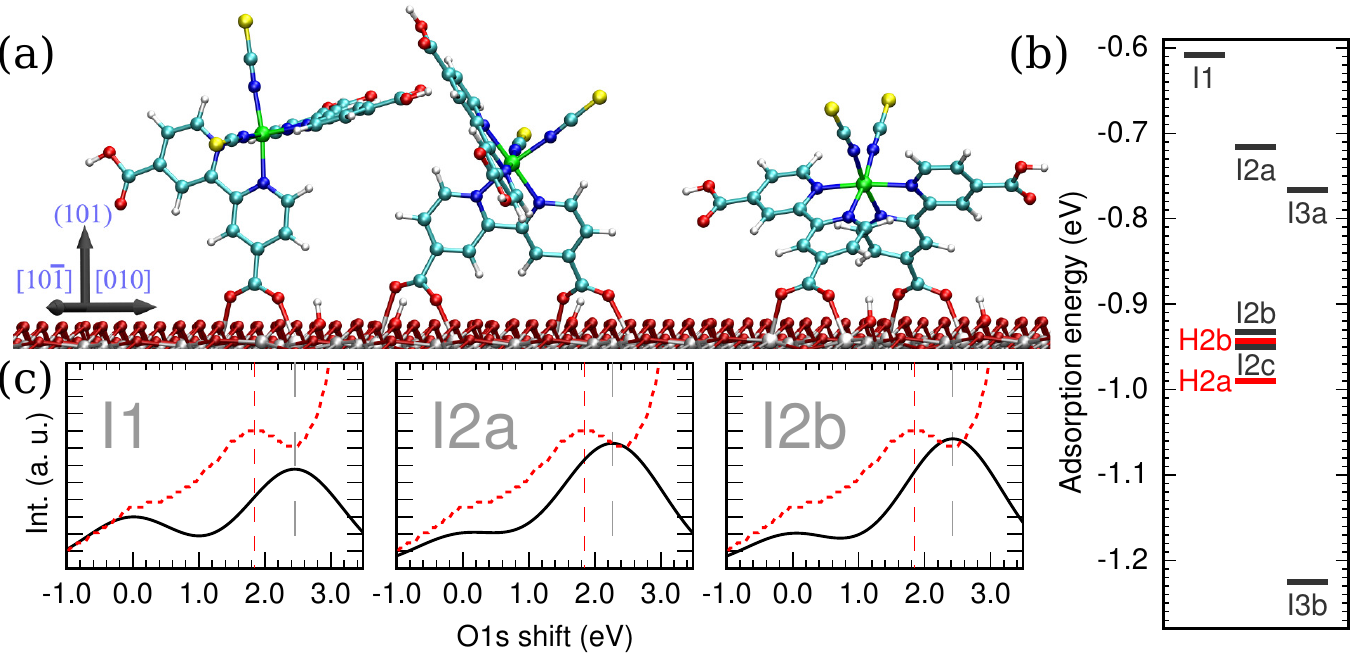}
\caption{\label{fig2}
(a)~Ball-and-stick representation of the interface models 
I1, I2a, and I2b.
Color code: Ru (green), S (yellow), N (blue), C (cyan),
O (red), H (white), Ti (silver).
(b)~Calculated adsorption energies for each interface model, with
stability increasing towards negative energies.
(c)~Calculated O$1s$ core-level spectra for the models
I1, I2a, and I2b (black solid lines) and experimental spectra from
Ref.~\onlinecite{Johansson2005} (red dashed lines). 
A Gaussian broadening of 1.6~eV, as estimated from
the spectra of Ref.~\onlinecite{Johansson2005}, has been applied to
the calculated spectra in order to account for core-hole lifetimes, 
vibrational broadening, and configurational disorder.
The peak arising from the TiO$_2$ substrate is not shown here but
can be seen in Fig.~\ref{fig3}(c).
All the spectra have been aligned to the leftmost peak.
The models I2c, I3a, and I3b and their calculated spectra are 
given as supplementary material \cite{sup_figs}, 
together with a quantitative analysis of all the spectral features.
}
\end{figure*}

Our ability to discriminate between candidate
interface models relies critically on accurate
core-level shift calculations.
In order to gauge the accuracy of the computational method
we considered a number of test molecules containing C and
O atoms whose structures are well understood.
In particular we included molecules which carry carboxylic acid groups
COOH similarly to the N3 dye. In Fig.~\ref{fig1}
we compare our calculated O$1s$ core-level shifts
with experiment \cite{Jolly1984,McQuaide1988}.
Our calculations exhibit very good agreement with experiment
over a wide energy range spanning 7~eV.
In the inset of Fig.~\ref{fig1} we concentrate on the molecules containing
carboxylic acid groups. In these groups the two oxygen atoms are inequivalent,
and the core electrons associated with the hydroxyl (COH) O atom
are more tightly bound than those associated with the carbonyl
(CO) O atom. Our calculations describe very accurately the differences
between the core-level shifts of the hydroxyl and of the carbonyl O atoms,
with an r.m.s.\ deviation from experiment below 0.2 eV.

For our model substrate we have chosen the anatase (101) surface,
which corresponds to the majority \cite{Lazzeri2001} of the total exposed surface
of the TiO$_2$ films used in DSCs and in Ref~\citenum{Johansson2005}. 
We considered eleven adsorption geometries of the N3 dye on this surface, 
including previously proposed models
\cite{Rensmo1999,Nazeeruddin2003,Schiffmann2010,Angelis2010}.
Schematic representations of these models can be seen in 
Figs.~\ref{fig2}(a),\ref{fig3}(a) and in the supplementary material \cite{sup_figs}.
Each model is labeled by the number 
of carboxylic groups which bind to the substrate.
The models I2a and I2b have been proposed in previous
experimental work \cite{Rensmo1999,Nazeeruddin2003},
and the models I2c and I3a have been introduced in
recent computational studies \cite{Schiffmann2010,Angelis2010}.

In order to make contact with previous studies
we report in Fig.~\ref{fig2}(b) the calculated adsorption energies 
for each interface \cite{sup_discussion}. 
The calculated adsorption energies span a range of 0.6~eV
across all the models considered. This range is comparable
to the energy of hydrogen bonds between
carboxylic acid groups in related systems. Indeed, the energy of the H-bond in the
formic acid dimer corresponds to 0.3 eV per monomer \cite{Tsuzuki2001}.
This observation suggests that hydrogen-bonding between the
carboxylic acid groups of N3 cannot be neglected
in the energetics of N3 adsorption on TiO$_2$.

In Fig.~\ref{fig2}(c) we compare our calculated O$1s$ core-level shifts 
\cite{sup_discussion} with the XPS measurements of Ref.~\onlinecite{Johansson2005}.
The measured spectra exhibit peaks at 529.8~eV, 531.4~eV and 533.2~eV.
The peak at the lowest binding energy (529.8~eV) has been
assigned to the O atoms of the TiO$_2$ substrate. The other two peaks have
been assigned to the inequivalent O atoms of the carboxylic groups 
in the dye \cite{Johansson2005}. Our calculations correctly reproduce 
the three measured peaks.
The uncertainty on the photoelectron escape depth \cite{Hufnerbook2},
surface stoichiometry, and H-coverage
makes the separation between the substrate peak at 529.8~eV and the two dye peaks 
unreliable for a quantitative comparison. 
We therefore concentrate on the dye peaks at 531.4~eV and at 533.2~eV. 
First we consider the intensity ratios of these peaks. 
The intensity of the peak at 533.2 eV scales
with the number of protonated carboxylic groups on the dye. The best match 
between our calculated intensities and experiment is obtained for 
interface models where the dye has two protonated carboxylic groups
(I2 and I3 models).
In model I1 the dye has three protonated COOH groups, 
leading to an intensity ratio (0.5) well off the experimental
estimate (0.3) \cite{Johansson2005}. We therefore reject the candidate 
model I1 on the grounds of intensity mismatch.
Second we consider the binding energy of the adsorbate peaks \cite{sup_discussion}.
As clearly shown in figure 2(c), the separations
of adsorbate peaks in all the
models of the I2 and I3 families fall within the range 2.3-2.6~eV, and
overestimate the measured peak separation of 1.8~eV \cite{Johansson2005}. 
This systematic deviation of 0.5-0.8~eV from experiment is well above our
0.2 eV error bar. We have carried out a number of tests in order to
confirm that such deviation is not a numerical artifact
\cite{sup_methods}.
We therefore assign the mismatch between theory and experiment 
to the inaccuracy of the models I2-I3.

By carrying out a detailed analysis of our calculated core-level shifts 
we noted that moderate changes in the structural parameters of the interface models,
such as dye twisting or bond length variations,
only lead to subtle changes in the shifts.
We therefore conclude that structural
variations across the models are not responsible for the observed 
0.5-0.8~eV deviation. 

These observations point us towards the possibility that supramolecular
interactions within the dye monolayer may play a role in the measured XPS spectra.
Our calculations for different surface coverages \cite{sup_discussion}
indicate that the separation between the dye peaks is not affected
by long-range electrostatic effects.
Hence sizeable changes in the calculated
peak separation can only arise from {\it short-range} interactions of the free
carboxylic groups in the dye with other molecules. Such interactions can 
happen in two ways: either some of the N3 carboxylic groups 
form bonds with contaminant molecules,
or the N3 dyes are bonded to each other within the monolayer.

The ex-situ preparation of the TiO$_2$/N3 interface of
Ref.~\onlinecite{Johansson2005} may lead to the presence of contaminant 
molecules in the system, such as water and hydrocarbons.
It is unlikely that large hydrocarbons systematically attach
to N3, but H$_2$O molecules are small 
enough to form hydrogen bonds with the COOH groups and may alter
the measured XPS spectra. 
However, our calculations of XPS spectra including water molecules
exhibit heavily distorted peak intensities \cite{sup_discussion},
and allow us to exclude this scenario on the grounds of intensity mismatch.

The only remaining possibility is that of dye-dye interactions
through the free COOH groups.
In order to test this hypothesis we considered two interface models,
H2a and H2b, which probe the limiting regimes of strong and weak H-bonding
respectively [Fig.~\ref{fig3}(a)]. Model H2a is derived from model I2a by forming
N3 dimers (H bond length 1.51 \AA). Model H2b is a self-assembled dye monolayer 
derived from model I2b (H bond length 2.68 \AA).
Figure 3b shows the XPS spectra calculated for the H-bonded
interface models. The agreement between theory and experiment
is seen to improve dramatically upon inclusion of supramolecular interactions
between the dyes. The calculated separation between dye 
peaks now matches the measured value within our error bar.
The modification of the XPS spectrum arising from supramolecular interactions
results from the lowering of the
O$1s$ binding energy in the hydroxyl group participating in the
hydrogen-bonding. This result is fully consistent
with previous experiments \cite{Joyner1976} and calculations
\cite{Aplincourt2001} on the formic acid dimer.
In model H2b this effect is less pronounced 
due to the weaker H-bond [Fig.~\ref{fig3}(b)].
Remarkably, if in the case of model H2a we include 
the substrate contribution to the spectrum \cite{sup_methods}, 
the agreement with experiment becomes excellent [Fig.~\ref{fig3}(c)].
These findings indicate that hydrogen-bonding 
between dyes is key to interpreting the photoemission data of 
Ref.~\onlinecite{Johansson2005}.

It is natural to ask whether additional H-bonded superstructures 
can exist at the TiO$_2$/N3 DSC interface. 
Elementary geometric considerations
show that, among all the model interfaces considered,
models H2a and H2b are the only possible H-bonded
homogeneous supramolecular structures \cite{sup_discussion}.
However, more complex heterogeneous assemblies 
of dyes cannot be excluded.

\begin{figure*}
\includegraphics{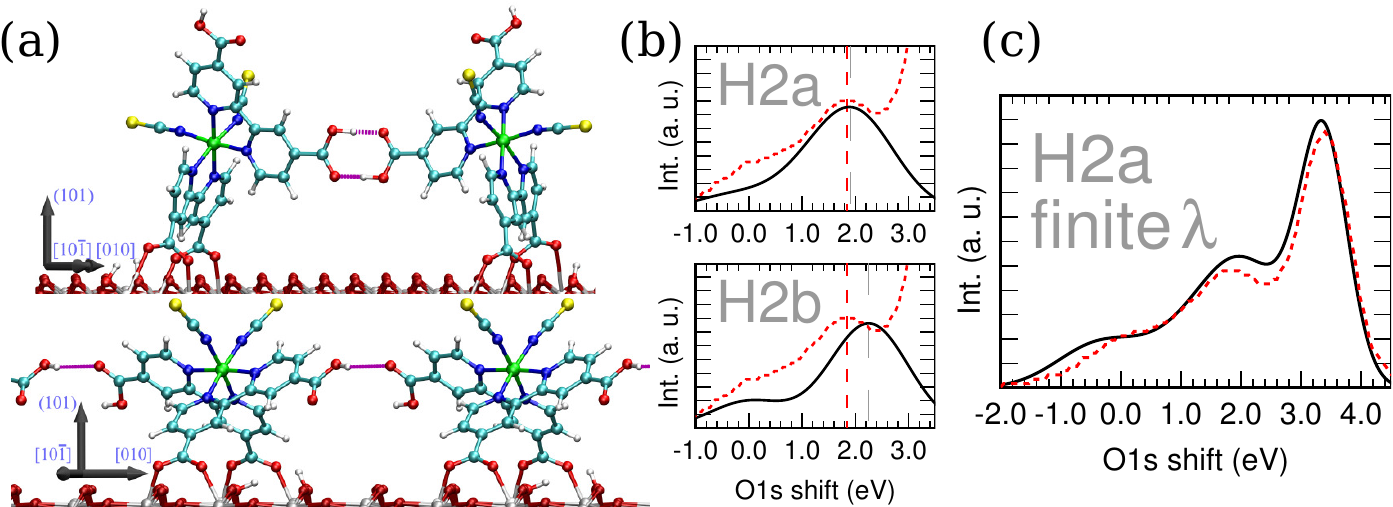}
\caption{\label{fig3}
(a)~Ball-and-stick representations of the interface models H2a and H2b.
(b)~Calculated O$1s$ core-level spectra for the models
H2a and H2b (black solid lines) and experimental spectra from
Ref.~\onlinecite{Johansson2005} (red dashed lines). 
(c)~Calculated O$1s$ core-level spectrum for the interface model H2a
including the contribution from the TiO$_2$ substrate and finite
escape depth effects
compared to the experimental spectrum.
}
\end{figure*}

STM experiments could directly probe the proposed H-bonded assembly.
Although there are reports of STM studies on anatase TiO$_2$ in the literature \cite{Gong2006}
to the best of our knowledge no data exists on N3-sensitized (101) surfaces.
However STM experiments of N3 on rutile TiO$_2$
reveal distinctively elongated features (ovals) in the tunneling maps \cite{Sasahara2010}. 
Figure \ref{fig3}(a) suggests that the dyes in our dimer model H2a
would naturally lead to an elongated STM footprint.
Since the calculated Ru-Ru distance of 1.6 nm in our model H2a
matches the length of the ovals in the STM maps (1.8 nm),
we speculate that the features observed in Ref.~\onlinecite{Sasahara2010} may
correspond to hydrogen-bonded N3 dimers.

It is worth asking whether our conclusions maintain their validity
for other important dyes.
The dye (Bu$_4$N)$_2$[Ru(dcbpyH)$_2$(NCS)$_2$] 
(N719) is structurally similar to N3, the only difference being that 
the protons on two carboxylic acid groups 
are replaced by counterions \cite{Nazeeruddin2003}.
Since in our interface model H2a the H-bonding does not occur
through the substituted groups, 
model H2a is also a possible candidate for the TiO$_2$/N719 interface.
Interestingly a very recent infrared and Raman study \cite{Lee2010} 
of the TiO$_2$/N719 
interface suggested that 
the dye may be involved in some form of hydrogen-bonding,
possibly with the substrate.
Our model H2a provides a natural explanation 
of the data of Ref.~\onlinecite{Lee2010} in terms of supramolecular
hydrogen-bonding.

In summary, we 
established that (i) models of the TiO$_2$/N3 interface based on isolated dyes
are unable to explain the measured XPS spectra, and (ii)
interface models where the N3 molecules form supramolecular hydrogen-bonded
assemblies are in good agreement with experiment.
Since the lowest photoexcited electronic state of N3 
is localized on the bipyridines and the carboxylic groups \cite{Fantacci2003},
we expect charge delocalization upon the formation of a supramolecular assembly,
with potential implications on the light absorption and the
electron transfer mechanisms in DSCs.
Our results are expected to hold also for other nanostructured solar cell
concepts, such as for instance solid-state DSCs \cite{Bach1998,Snaith2007}.
The present finding highlights the importance of supramolecular
interactions at semiconductor/dye interfaces,
and bears implications for the design of more efficient nanostructured solar cells.

We thank F. De Angelis and H. Snaith for fruitful
discussions. This work is supported by
the UK EPSRC and the ERC under the EU FP7 / ERC grant no. 239578.
Calculations were performed in part at the 
Oxford Supercomputing Centre. Figures rendered using VMD \cite{VMD}.


%


\section*{Supplementary material}

10

# Hydrogen-bonded supramolecular assembly of dyes

# at nanostructured solar cell interfaces:

# Supplementary Information

Christopher E. Patrick and Feliciano Giustino

*Department of Materials, University of Oxford,*

*Parks Road, Oxford OX1 3PH, United Kingdom*

The following Supplementary Information is provided.

**Computational Methods**

**Computational Methods**

The calculations were performed using density functional theory (DFT) within the generalized gradient approximation of Ref. 1. We used periodic simulation cells and described the electronic wavefunctions and charge density using plane wave basis sets as implemented in the `Quantum ESPRESSO` software distribution [2]. The core-valence interaction was taken into account by means of ultrasoft pseudopotentials [3]. The structures were relaxed via damped Car-Parrinello molecular dynamics by sampling the Brillouin zone at the $\Gamma$ point [4, 5]. In order to generate the substrate model we optimized the bulk anatase $TiO_2$ lattice parameters by sampling the Brillouin zone on six inequivalent Monkhorst-Pack points while keeping the $I4_1/amd$ symmetry fixed. We constructed a stoichiometric slab by taking a cut through the bulk $TiO_2$ anatase such that the (101) surface was exposed. We described the $TiO_2$ surface using a rectangular slab of area $20.9 \times 19.0$ Å$^2$ for the six interface models I1-I3. In the case of the H-bonded configurations we used a periodic rectangular slab with area $31.3 \times 7.6$ Å$^2$ for model H2a, and an oblique cell with area 138 Å$^2$ for model H2b. For all interface models the thickness of the $TiO_2$ slabs was fixed to 12 layers of atoms (5.8 Å) and the interaction between periodic replicas along the direction perpendicular to the surface was minimized by including a vacuum region of 10 Å. In order to compensate for the electrostatic interactions between the interface replicas we calculated the core-level shifts including self-consistently the dipole correction of Ref. 6. During the geometry optimizations the bottom three layers of the slab were fixed in their bulk positions in order to mimic the structure of the $TiO_2$ nanoparticle far from the surface. Structural relaxations were carried out until the force on each atom was below 0.07 eV/Å. Calculations were found to be converged with kinetic energy cutoffs of 35 Ry and 200 Ry for the electron wavefunctions and charge density, respectively.

O1$s$ core-level shifts were calculated by following the method of Refs. 7, 8 which takes into account final state effects. In the calculations with a core hole charge neutrality was restored by using a positive jellium background. For the gas phase molecules we calculated the core-level shifts for various cell sizes and used the Makov-Payne expansion [9] for the extrapolation to infinite simulation cells. Due to error cancellation the difference between the O1$s$ shifts of two O atoms in the same computational cell converges much faster with increasing cell size as compared to the individual shifts. For example, in the case of formic acid on changing the

cell size from $9 \times 9 \times 9$ Å$^3$ to $27 \times 27 \times 27$ Å$^3$ the individual shifts of the O atoms change by 0.25 eV. However, the difference between the shifts of the carbonyl O atom and the hydroxyl O atom change by less than 0.02 eV. For this reason in the interface calculations we used simulation cells identical to those adopted for the geometry optimizations. Core-level shift calculations for the interface models using thicker slabs of 24 atomic layers did not yield any significant differences with respect to the 12-layer calculations.

There is an uncertainty in the calculation of the substrate O1$s$ peak at 529.8 eV. In fact the number of TiO$_2$ layers which contribute to this peak depends on the photoelectron escape depth. In addition surface dipoles associated with possible surface defects may affect the energy separation between the substrate and the two adsorbate peaks at 531.4 eV and 533.2 eV. In the calculation of the full spectrum in Figure 3(c) we only included the topmost layer of O atoms, and we used an escape depth of 10 Å. The latter value has been estimated from the inelastic mean free paths reported in Ref. 10 for a photon energy of 758 eV [11].

Since DFT might not describe hydrogen bonds accurately, we conducted extensive tests on the geometry and core-level shifts of H-bonded systems. For example we calculated the hydrogen-bond energy in the formic acid dimer to be 0.38 eV per molecule. This value compares favorably with the value of 0.3 eV from the MP2 calculations and spectroscopic data reported in Ref. 12. The difference between the core-level shifts of the carbonyl O atom and the hydroxyl O atom was calculated to be 1.19 eV. This value compares favorably with experiment, yielding 1.3 eV [13].

The broadening of the photoemission data of Ref. 11 arises from the finite lifetimes of the core-holes, from vibrational broadening, and from the averaging over all the possible adsorption configurations. In the present study we do not address these aspects. In particular our best candidate interface models are meant to describe only the dye adsorption configuration with the highest yield.

While the present study focuses on XPS experiments performed on dry interfaces (i.e. without the redox electrolyte), our conclusions are expected to remain valid even for complete DSC devices because the electrolyte is introduced after the sensitization step, when the TiO$_2$/N3 interface has already formed. The present study also bear relevance to solid-state DSCs where the electrolyte is replaced by a molecular hole-transporter [14, 15].

4

**Supplementary Note 1:**

The hydrogen-bonded adsorption models considered in the main text are derived from the single molecule adsorption modes I2a and I2b. Model H2a is obtained from model I2a by forming hydrogen bonds between two dyes (bond length 1.51 Å). Model H2b is a hydrogen-bonded chain of dyes derived from model I2b (bond length 2.68 Å). The formation of hydrogen bonds stabilizes the interface in both models, by 270 meV per dye in model H2a, and by 20 meV per dye in model H2b [Figure 2(b)].

H-bonded dye chains were also proposed in Ref. 16 but in that work one of the two H-bonded carboxylic groups is deprotonated, while the same carboxylic groups are fully protonated in our model H2b. We tested the interface model proposed in Ref. 16 and found that the loss of a proton halves the intensity of the dye peak at 533.2 eV, resulting in a calculated core-level spectrum in sharp disagreement with experiment. Incidentally we note that we calculate the model proposed in Ref. 16 to be considerably less stable than all the other models considered here.

**Supplementary Note 2:**

While a dimer of model I2b could be formed by mixing different enantiomers of the N3 dye, there would be a mismatch between the positions of the anchor carboxylic groups and the Ti chemisorption sites on the $TiO_2$ surface. The formation of strong H-bonds within a monolayer of dyes adsorbed as in the interface models I3a, I3b and I2c is prevented by the unadsorbed carboxylic group pointing out of the layer. The interface model I1 is rejected on grounds of intensity mismatch.

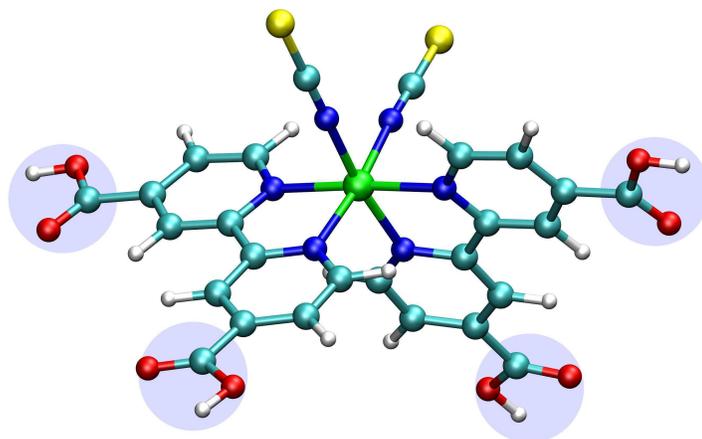

FIG. 1. The N3 molecule

A central Ru atom is sixfold coordinated to the N atoms of two bipyridines and two thiocyanate ligands, with each bipyridine carrying two carboxylic acid groups (highlighted). The carboxylic acid groups anchor the molecule to the substrate via Ti-O bonds.

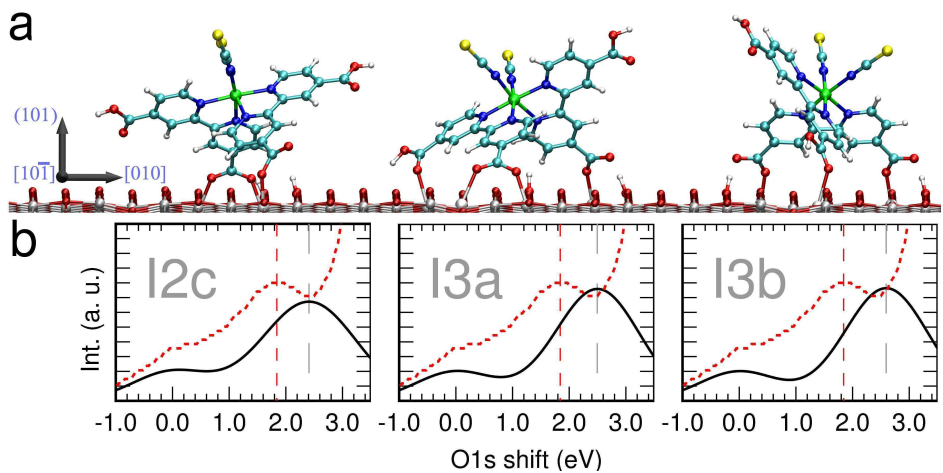

FIG. 2. Additional single-molecule adsorption models considered

(a) Ball-and-stick representations of interface models I2c, I3a, and I3b. In interface models I2a and I2b [Fig. 2(a) in main text] the dye binds to the TiO$_2$ substrate via two carboxylic groups, both in bridging bidentate modes. In both models the H atoms from the binding carboxylic groups bind to the substrate O atoms [17]. In model I2a the bridging carboxylic groups belong to the same bipyridine [18], while in model I2b the carboxylic groups belong to different bipyridines [19]. In model I2c, the two carboxylic groups which participate in binding belong to different bipyridines, with one in a bridging mode and the other in a unidentate mode. This configuration was proposed in Ref. 16. In model I3a the N3 dye binds to the substrate via one bridging carboxylic group and two unidentate groups. This configuration was proposed in Ref. 20 for the related N719 dye. In model I3b the three carboxylic groups bind to the substrate in unidentate modes. In the three interface models I2c, I3a and I3b the interaction of the dye with the protons on the TiO$_2$ surface stabilizes the structure, consistent with studies of formic acid on the same surface [17]. Model I3b is the most energetically favourable owing to the minimum strain exerted on the dye molecule.

(b) Calculated O1$s$ core-level spectra for the interface models I2c, I3a, and I3b (black solid line), compared to the experimental spectrum of Ref. 11 (red dashed line). The calculated spectra systematically overestimate the separation of the peaks as obtained from experiment, as indicated by the vertical dashed lines. The peak energies and intensities are reported in Supplementary Table 2. We note from both here and Figure 2(b) in the main text that substantial changes in structural parameters lead to subtle changes in shifts. For instance, the spectrum of model I2c is remarkably similar to that of I2a and I2b, even though the dye binds to the substrate through 4 Ti-O bonds in models I2a and I2b and only 3 Ti-O bonds in I2c. It is also interesting to note that the disagreement between theory and experiment is most severe for model I3b, which is calculated to be the most energetically stable configuration [Figure 2(b) in the main text].

7

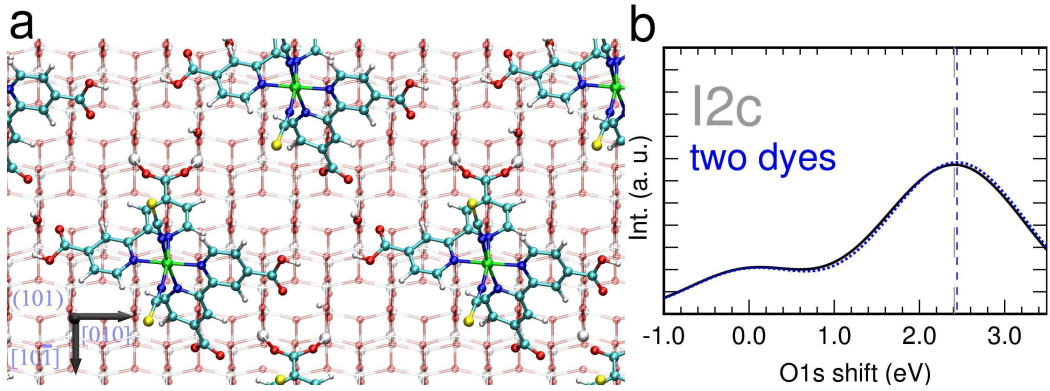

FIG. 3. Effect of increased dye coverage on O1s spectra

(a) Ball-and-stick representation of the interface model I2c with two dyes per simulation cell. In this new configuration the areal density is 0.5 nm$^{-2}$ [compared to 0.25 nm$^{-2}$ for the data presented in Supplementary Figure 2(b)]. The shortest distance between the carboxylic groups of neighbouring dyes is 4.4 Å. Consequently no hydrogen bonds are formed between neighbouring dyes.

(b) Calculated O1s core-level spectrum for the model interface with higher surface coverage (blue dotted line), compared to the calculated spectrum of model I2c (black line). The peak energies in the two spectra differ by less than 0.03 eV. We conclude that the observed discrepancy in peak separation in the experimental and calculated spectra is not explained by long-range electrostatic effects.

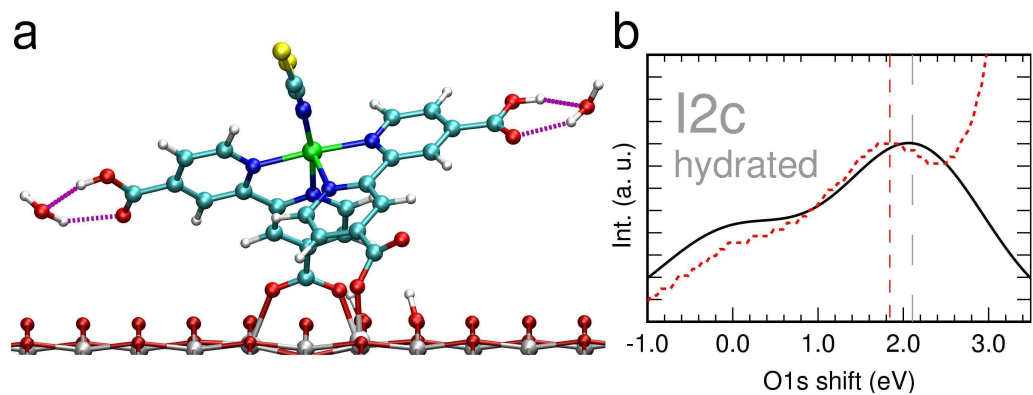

FIG. 4. Effect of $H_2O$ on $O1s$ spectra

(a) Ball-and-stick representation of the interface model I2c with additional water molecules forming H-bonds with each carboxylic acid group in the dye. The bond lengths of the H-bonds formed with the hydroxyl group and with the carbonyl group are 1.80 Å and 2.05 Å respectively.

(b) Calculated $O1s$ core-level spectrum of this interface model. The H-bonding reduces the separation between the two dye peaks to 2.1 eV, leading to a good agreement with the experimental separation of 1.9 eV measured in Ref. 11 (red dashed line). However, the O atom of the $H_2O$ molecule contributes to the leftmost peak. The resulting intensity ratio of 0.5 is in disagreement with the experimental value of 0.3. On taking into account the finite escape depth of the photoelectrons the disagreement becomes even more pronounced. In general, if other contaminant molecules were to alter the measured XPS spectra, their areal density would have to be comparable to the dye surface coverage, and additional features would appear in the measured XPS spectra. This scenario is in contrast with the findings of Ref. 11.

TABLE I. O1$s$ shifts of test molecules. All values are referenced to the shift of the H$_2$O molecule. For molecules containing carboxylic acid groups the shifts of the carbonyl O atoms and of the hydroxyl O atoms are reported separately. In the other cases where a molecule has more than one O atom, the relevant atom is indicated in boldface. The experimental data are from Refs. 21, 22. We have also calculated the separation between carbonyl and hydroxyl O1$s$ core-level shifts in N3 (2.5 eV), isonicotinic acid (2.1 eV) and bi-isonicotinic acid (2.2 eV) but we are unaware of any published gas phase data for these low volatility molecules.

| Molecule | Theory (eV) | Experiment (eV) |
| --- | --- | --- |
| H$_2$O | 0.00 | 0.00 |
| O$_2$ | -4.18 | -3.8 |
| N$_2$**O** | -1.82 | -1.5 |
| F$_3$C**OO**OCF$_3$ (average) | -1.83 | -1.6 |
| F$_3$CO**O**OCF$_3$ | -2.94 | -2.8 |
| N-(OH)-2-Pyridone (CO) | 2.85 | 2.9 |
| N-(OH)-2-Pyridone (OH) | 0.09 | -0.2 |
| CF$_3$N**O** | -2.12 | -2.4 |
| HCOOH (Formic acid) CO | 0.99 | 0.9 |
| HCOOH OH | -0.80 | -0.7 |
| CH$_3$COOH (Acetic acid) CO | 1.64 | 1.6 |
| CH$_3$COOH OH | -0.25 | -0.2 |
| CH$_3$CH$_2$COOH (Propionic acid) CO | 1.90 | 1.6 |
| CH$_3$CH$_2$COOH OH | 0.01 | -0.1 |
| C$_6$H$_5$OH (Phenol) | 0.76 | 1.0 |
| F$_2$CHCOOH (Difluoroacetic acid) CO | 0.78 | 0.6 |
| F$_2$CHCOOH OH | -1.12 | -1.1 |
| CF$_3$COOH (Trifluoroacetic acid) CO | 0.42 | 0.3 |
| CF$_3$COOH OH | -1.45 | -1.4 |
| C$_6$H$_5$COOH (Benzoic acid) CO | 2.42 | 2.2 |
| C$_6$H$_5$COOH OH | 0.30 | 0.1 |
| C$_6$H$_4$(COOH)$_2$ (Phthalic acid) CO | 2.51 | 1.8 |
| C$_6$H$_4$(COOH)$_2$ OH | 0.42 | -0.1 |
| C$_6$H$_4$(COOH)$_2$ (Isophthalic acid) CO | 2.28 | 1.9 |
| C$_6$H$_4$(COOH)$_2$ OH | 0.10 | -0.2 |
| C$_6$H$_4$(COOH)$_2$ (Terephthalic acid) CO | 2.28 | 1.8 |
| C$_6$H$_4$(COOH)$_2$ OH | 0.02 | -0.2 |

TABLE II. Calculated energy separation and intensity ratio between the two dye peaks for each interface model considered, compared to the experimental data of Ref. 11. The calculations were performed in the limit of infinite escape depth.

| Model | separation (eV) | intensity ratio |
|---|---|---|
| I1 | 2.45 | 0.50 |
| I2a | 2.25 | 0.25 |
| I2b | 2.41 | 0.25 |
| I2c | 2.41 | 0.25 |
| I3a | 2.48 | 0.25 |
| I3b | 2.59 | 0.25 |
| H2a | 1.92 | 0.21 |
| H2b | 2.24 | 0.25 |
| Experiment | 1.8 | 0.31 |



\end{document}